\documentclass[conference]{IEEEtran}
\usepackage{hyperref}
\IEEEoverridecommandlockouts
\usepackage{cite}
\usepackage{listings}
\usepackage{xcolor}
\usepackage{amsmath,amssymb,amsfonts}
\usepackage{algorithmic}
\usepackage{graphicx}
\usepackage{booktabs}
\usepackage{array}
\usepackage{textcomp}
\usepackage{xcolor}
\usepackage{float}
\usepackage{placeins}
\def\BibTeX{{\rm B\kern-.05em{\sc i\kern-.025em b}\kern-.08em
    T\kern-.1667em\lower.7ex\hbox{E}\kern-.125emX}}
    \makeatletter
\newcommand{\linebreakand}{%
  \end{@IEEEauthorhalign}
  \hfill\mbox{}\par
  \mbox{}\hfill\begin{@IEEEauthorhalign}
}
\makeatother
\begin{document}

\title{IndicEval-XL: Bridging Linguistic Diversity in Code Generation Across Indic Languages \\
\thanks{Find dataset at \url{https://github.com/telekom/IndicEval-XL}}
}

\author{\IEEEauthorblockN{Ujjwal Singh}
\IEEEauthorblockA{\textit{Deutsche Telekom Digital Labs} \\
Gurugram, India \\
ujjwal.singh@telekom-digital.com}
\and
\IEEEauthorblockN{Aditi Sharma}
\IEEEauthorblockA{\textit{Deutsche Telekom Digital labs} \\
Gurugram, India \\
aditi.sharma1@telekom-digital.com}
\and
\IEEEauthorblockN{Nikhil Gupta}
\IEEEauthorblockA{\textit{Deutsche Telekom Digital Labs} \\
Gurugram, India \\
nikhil@telekom-digital.com}
\linebreakand
\IEEEauthorblockN{Deepakshi}
\IEEEauthorblockA{\textit{Deutsche Telekom Digital Labs} \\
Gurugram, India \\
deepakshi.ext1@telekom-digital.com}
\and
\IEEEauthorblockN{Vivek Kumar Jha}
\IEEEauthorblockA{\textit{Deutsche Telekom Digital Labs} \\
Gurugram, India \\
vivek.jha@telekom-digital.com}
}

\maketitle

\begin{abstract}
Large Language Models (LLMs) have demonstrated remarkable capabilities in code generation from natural language prompts, revolutionizing software development workflows. As we advance towards agent-based development paradigms, these models form the cornerstone of next-generation software development lifecycles. However, current benchmarks for evaluating multilingual code generation capabilities are predominantly English-centric, limiting their applicability across the global developer community.
To address this limitation, we present IndicEval-XL, a comprehensive benchmark for code generation that incorporates 6 major Indic languages, collectively spoken by approximately 14\% of the world's population. Our benchmark bridges these languages with 12 programming languages, creating a robust evaluation framework. This work is particularly significant given India's representation of one-eighth of the global population and the crucial role Indic languages play in Indian society. 
IndicEval-XL represents a significant step toward expanding the linguistic diversity in code generation systems and evaluation frameworks. By developing resources that support multiple languages, we aim to make AI-powered development tools more inclusive and accessible to developers of various linguistic backgrounds. To facilitate further research and development in this direction, we make our dataset and evaluation benchmark publicly available at \href{https://github.com/telekom/IndicEval-XL}{github}. \newline
\end{abstract}

\section{Introduction}
Modern code generation merges computational linguistics and artificial intelligence to convert human intentions, expressed in natural language, into executable programming code using probabilistic, symbolic, and neural strategies. Early efforts featured probabilistic context-free grammars, leveraging abstract syntax trees (ASTs) to enforce hierarchical syntax constraints and accommodate variations in user instructions \cite{zettlemoyer2005learning}, \cite{yin2017syntactic}. This AST-centric perspective moved beyond rule-based approaches by incorporating probabilistic elements that capture programming nuances in a structured representation.

\textbf{Transition to Neural Architectures}
Subsequent work integrated neural sequence modeling with external memory, as shown in the Neural Turing Machine and the Differentiable Neural Computer, enabling iterative and adaptive reasoning for more complex program synthesis tasks \cite{graves2014neural}, \cite{graves2016hybrid}. Later expansions, such as the Neural Program Interpreter, incorporated recurrence mechanisms to manage sequential dependencies across multi-step solutions, allowing deeper learning of execution pathways \cite{reed2015neural}. This shift significantly broadened the capabilities of program synthesis, facilitating models that automatically learn to map textual prompts to functionally correct code across diverse problem sets.
\newline
\newline
\textbf{Evaluation and Benchmarking}
Token-level metrics such as BLEU provided initial insights but often failed to capture functional correctness or behavioral validity \cite{papineni2002bleu}. The introduction of the pass@k metric underscored the necessity of validating generated codes using test suites, emphasizing the ratio of solutions that pass real-world standards within a fixed number of attempts \cite{chen2021evaluating}. Datasets such as HumanEval and APPS focused on natural language-to-code tasks, containing programming challenges of varying complexities and comprehensive test cases for more precise evaluation \cite{hendrycks2021measuring}, \cite{chen2021evaluating}. HumanEval particularly highlighted functional correctness, while APPS presented competitive programming tasks, testing adaptability and generalization in complex problem domains.
\newline
\newline
\begin{figure*}[t]
  \centering
  \includegraphics[width=1\textwidth]{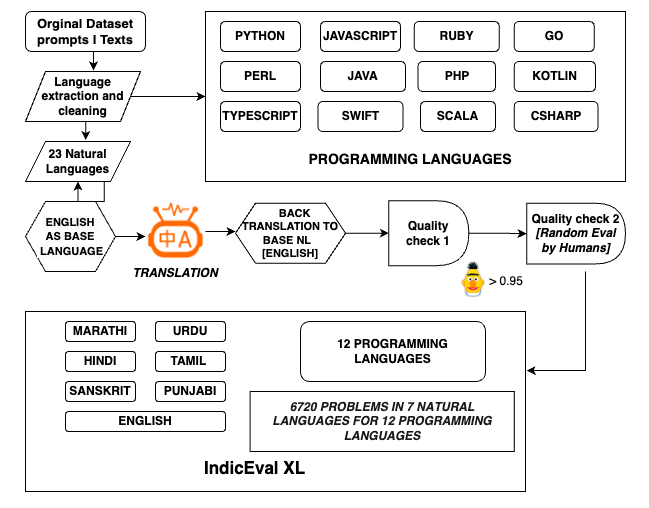} 
  \caption{IndicEval-XL Data Creation Illustration. This diagram explains end to end process of creating and Validating the IndicEval-XL dataset, across all the 6 Indic Languages and English. In quality check 1, along with bertscore (major) we have also done minor checks using \textbf{(BLEU $>$ 25)} and \textbf{(METEOR $>$ 0.5)} for the back translated text (Using GPT-4) for quality verification.}
\end{figure*}
\textbf{Multilingual and Cross-Lingual Expansions}
Recent advances underscore the importance of multilingual code generation, exemplified by HumanEval-XL’s inclusion of 23 natural languages and 12 programming languages, illustrating robust cross-lingual alignment methods . Parallel efforts have targeted underrepresented languages, notably those of India, through resources like IndicNLP and MuRIL, which address the morphological diversity and syntactic complexity inherent to Indian languages \cite{kakwani2020indicnlp}, \cite{khanuja2021muril}. The growing emphasis on linguistic inclusivity has led to datasets designed for Hindi, English, Marathi, Punjabi, Sanskrit, and Tamil, capturing nuances ranging from Sanskrit’s inflectional constructs to Tamil’s agglutinative forms for more expansive global applicability. By mirroring the growing diversity of programming needs and linguistic contexts, these multilingual benchmarks aim to foster inclusive AI systems that maintain functional precision and adapt to a wide spectrum of syntactic constraints.
\newline
\newline
Overall, the synergy between structured representations, neural architectures, and cross-lingual benchmarks has pushed code generation beyond template-based outputs toward robust, context-aware program synthesis \cite{sharma2022developing}. Continuing research centers on refining multilingual capabilities, advancing evaluation metrics, and constructing inclusive datasets, ensuring that the next generation of program synthesis models can effectively handle diverse languages and programming paradigms.

\section{IndicEval - XL}
We introduce \textbf{IndicEval-XL} benchmark, where our primary objective is to provide an accessible code generation benchmark, focusing on major Indic languages spoken by over 14 percent of the world population. In preparation for this dataset, we have adhered to the following guiding principles to ensure the dataset is of the utmost quality.

\subsection{Methodology}

We focused on picking up the task, that is complex in nature. LLMs have the profound potential to impact the way we code today, thus requiring a robust benchmarking . To access the code generation capabilities of the LLMs, we employ BERTScore and CodeBERTScore evalution metrics \cite{zhang2020bertscore},\cite{pan2021codebert}. We do not  
use traditional pass@k since we saw smaller LLM struggle to keep the function names correct as per the test cases, even though the actual function is correct \cite{smith2020neural}. In one case, the solution can be done in both recursive and iterative ways, but the test cases were written in iterative ways only, which caused the provision of fundamentally wrong metrics. We have  discussed in detail why this is the case in experimentation section. In our dataset, we realize that  
pass@k is still valuable for some use cases in LLMs, so we have kept the manual written test cases as well.

\subsection{Diversity in dataset}
Diversity should be at the heart of any technological advancements, especially in case of AI where systems have potential to become smarter than humans\cite{gemini2025}. We have carefully curated Indic languages that are spoken by more than 14\% of the world population. Since most of the code-eval benchmarks focus only on the popular languages (English) dominant, we tried to bridge this gap with our work. This ensures fair and insighful LLM evaluations across the various linguistic diversification.

\subsection{Accessibility}
To ensure that IndicEval - XL is free to use and accessible to use for the true benchmarking of LLMs along with other datasets and benchmarks, all the content used in this data set is free to use and redistribute in a completely open and unrestricted manner for research purposes.

\section{Dataset Construction}
In this work, we are introducing a robust way to increase the benchmarking standards of LLMs in multilingual code generation paradigm, bridging the liguistic gap. We took an approach of finding NL-PL pairs of problems, and extended it for various popular indic languages . In this work, we provide two sets of datasets \textbf{HumanEval - XXL (multilingual code generation dataset)} which comprises of 29 NL-PL pairs of problems (22 from HumanEval-XL \cite{peng2024humanevalxlmultilingualcodegeneration} and 7 from IndicEval-XL) and \textbf{IndicEval-XL} which primarily focuses on 6 Indic Languages. We have made both our datasets publicly available at \url{https://github.com/telekom/IndicEval-XL}. In Fig.1 we have described how this dataset is constructed.

\subsection{Data curation and Pre-Processing}
We started with the HumanEval XL dataset \cite{peng2024humanevalxlmultilingualcodegeneration}, we extracted the text prompts, and we separated out these text prompts for all the 12 programming languages. We took English as the base language, since in the base dataset, English is the one most curated. We created 12 NL-PL where the NL is English language; this gives us flexibility to convert these English prompts into various Indian languages.

\subsection{Translations}
We tried two different approaches for translation, the first is direct translation with LLMs like GPT-4o\cite{flashapi2024}. What we observed is even though they are remarkably accurate in Hindi, for non-English languages like Sanskrit, they were not working as expected. We found that there might be a huge gap in the base data that is used for compression for certain languages. We devised a second more robust and cost effective approach is to use specially trained transformers models like IndicTrans2 \cite{gala2023indictrans2}. They are relatively cheap (1B paramters) model and provide much better results in our case. We passed this approach with rigorous quality checks in the later stages\cite{comparative2023}.

\subsection{Quality Checks}
We employ robust checks to ensure that the data presented in this work are of the utmost quality. To ensure this, we have taken our translated results and translated them back to the original languages (English). After this, we take these back translated examples and pass them through the BLEU and METEOR benchmarks\cite{papineni2002bleu},\cite{banerjee2005meteor}, we set the bar for BLUE to be in the range of 25-30 and Meteor at $>$ 0.25. Almost all of the samples passed this stage. The ones that did not, we corrected them manually. Then we passed them through the second level of quality check, where we used BERTScore with RoBERTa \cite{zhang2020bertscore},\cite{liu2019roberta}. We kept bar very high in this case, any sample whose similarity score between original text and back translated text is less than 0.95, we discarded it and manually improved them.

\subsection{Quality Control}
To further strengthen the quality of the data set, after quality checks, we randomly selected 10-12 data points from each data-set of languages and manually evaluated them to understand the contextual inferences of these data samples, and if the translated text in these languages captures the essence of the original prompt texts or not. After all the checks we present \textbf{IndicEval-XL} benchmark, comprising of 80 parallel programming languages problem for 6 Indic Languages and English Language (7 NL in total). IndicEval-XL contains 6,720 coding problems. Percentages are calculated using the 2025 world population estimate of 8 billion .
\begin{table}[ht]
\centering
\caption{Global Distribution of Selected Indic Languages (2025 Estimates)}
\label{tab:lang_dist}
\renewcommand{\arraystretch}{1.2} 
\setlength{\tabcolsep}{3pt} 
\begin{tabular}{p{1.2cm}p{1.8cm}p{4.2cm}}
\toprule
\textbf{Language} & \textbf{World Population \%} & \textbf{Primary Geographic Locations} \\
\midrule
Hindi & 7.61\% & India, Nepal, Mauritius, Fiji, South Africa, USA \\
Sanskrit & 0.0003\% & India, Germany, USA, UK, Nepal \\
Punjabi & 1.88\% & Pakistan, India, Canada, UK, USA, Australia \\
Urdu & 2.98\% & Pakistan, India, Saudi Arabia, UAE, UK \\
Tamil & 1.08\% & India, Sri Lanka, Singapore, Malaysia \\
Marathi & 1.19\% & India (Maharashtra, Gujarat, Karnataka) \\
\bottomrule
\end{tabular}
\vspace{-0.5em}
\end{table}
\small
\textit Percentages calculated using 2025 world population estimate of 8 billion. Sanskrit figures represent self-reported native speakers; functional speakers may be higher due to liturgical/educational use. Urdu and Hindi share partial mutual intelligibility but are counted separately here.
\newline
\newline
\textbf{Overview of Dataset} This dataset has 12 Programming Languages and 7 Natural Languages. 12 PLs are same as present in multilingual HumanEval Languages i.e. Python, Java, Go, Kotlin, PHP, Ruby, Scala, JavaScript, C\#, Perl, Swift and TypeScript. 7 NLs includes English and 6 Indic Languages spoken across the Indian Peninsular and world by more than 14$\%$ of the world population, as depicted in Table 1. We have included sanskrit in our dataset creation as it is one of the most ancient languages of India, and we want to explore how well LLMs can perform in these ancient languages with very less data available in the Internet\cite{comparative2023}.

\section{Experiments}
In our evaluation, we focused on both large language models (LLMs) and small language models . For LLMs, we selected \textbf{Gemini 1.5} and \textbf{Gemini 2.0 Flash Thinking}\cite{gemini2025v2},\cite{fellowai2024gemini}, while for SLMs, we chose \textbf{LLaMA 7B}\cite{meta2024llama7b}. 
These models demonstrate state-of-the-art performance within their respective categories. To evaluate their performance, we have primarily employed CodeBERTScore \cite{pan2021codebert} rather than the pass@k metric\cite{lyu2024passimprovecodegeneration}, given inherent limitations in the latter. The pass@k benchmark assesses generated code based on its ability to pass predefined human-written test cases. However, our analysis revealed that small language models (SLMs) often fail due to minor syntactic discrepancies. For example, even when presented with a semantically correct implementation, a model could misname a function or adopt an alternate coding strategy, such as using recursion instead of an iterative approach, when the corresponding test cases are designed for the iterative version. As illustrated in Table 2, even though the code generated by LLaMA is functionally correct, the variation in the naming conventions result in a zero score under the pass@k metric. This rigid evaluation framework unduly penalizes SLMs and does not fully capture their ability to produce functionally accurate solutions, leading to an underestimation of their true performance.

\begin{figure*}[t]
\centering
\includegraphics[width=\textwidth, height=1\textwidth, keepaspectratio]{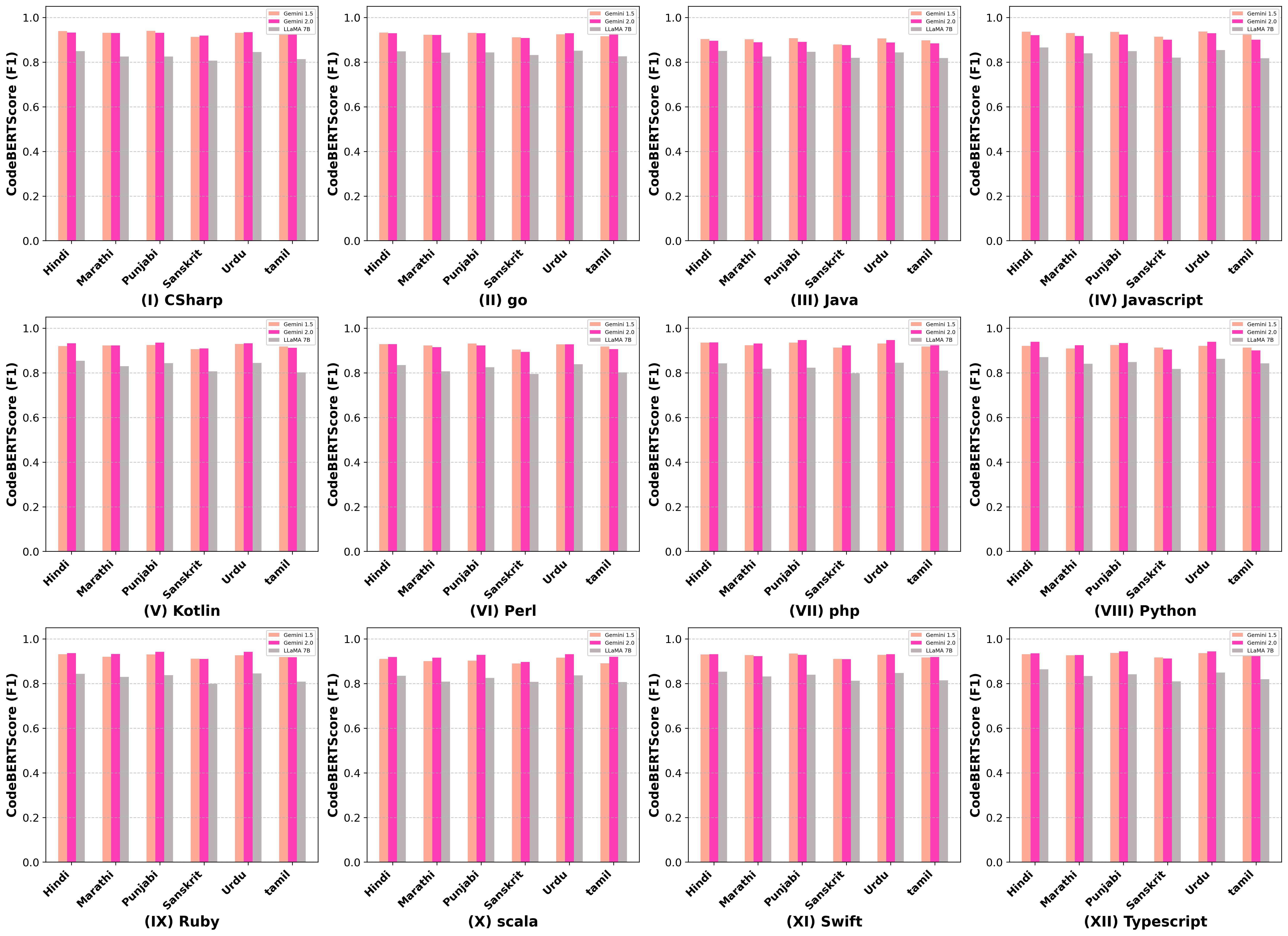}
\caption{The above figure shows the codeBERTScore of each of the 6 Indic languages across 12 programming languages. These performance scores are of 3 LLMs i.e. Gemini 1.5, Gemini 2.0 Flash Thinking and Llama 7b. Languages are ordered randonly in the given graph.}
\label{fig:codebert}
\end{figure*}

\subsection{CodeBERTScore}
CodeBERTScore is a crucial metric for evaluating code generation models, especially in multilingual environments. Unlike traditional metrics such as BLEU and ROUGE, which
focus on token-level matches, CodeBERTScore leverages contextual embeddings to
assess semantic similarity, ensuring that the generated code maintains functional
correctness across different languages. This is particularly beneficial for Indic
languages, where syntactic variations do not necessarily alter a program’s logic. By
computing the cosine similarity between the candidate and the reference code embeddings,
CodeBERTScore provides a more accurate representation of semantic preservation.

\begin{table}[ht]
\centering
\caption{The is\_prime\_function generated by LLaMA 7B is semantically correct but fails all test cases due to naming convention issues.}
\label{tab:vert_images}
\lstset{basicstyle=\footnotesize\ttfamily, breaklines=true}
\begin{lstlisting}
class Solution { 
/** 
* Checks if a given number is prime. 
* @param n The number to check. 
* @return True if the number is prime, false otherwise. 
*/
   public boolean is_prime_number(int n) { 
      if (n <= 1) { 
        return false; 
       } 
      for (int i = 2; ii <= n; i++) { 
          if (n % i == 0) { 
            return false; 
        } 
      } 
      return true; 
   }
}
\end{lstlisting}

\begin{lstlisting}
class Main { 
   public static boolean compare(Object obj1, Object obj2) { 
      if (obj1==null && obj2==null){ 
         return true; 
      } else if (obj1 == null || obj2 == null){ 
         return false; 
      } else { 
         return obj1.equals(obj2); 
      } 
   } 
   public static void main(String[] args) throws Exception { 
     int arg00 = 6; 
     Boolean x0 = IsPrime.isPrime(6); 
     Boolean v0= false; 
     if (!(compare(x0, v0))) { 
        throw new java.lang.Exception("Exception test case 0 did not pass. x0 = " + x0);} 
     int arg10 = 101; 
     Boolean x1 = IsPrime.isPrime(101); 
     Boolean v1 = true; 
     if (!(compare(x1, v1))) { 
        throw new java.lang.Exception("Exception -- test case 1 did not pass. x1 =" + x1); } 
     int arg20 = 11; 
     Boolean x2 = IsPrime.isPrime(11); 
     Boolean v2 = true; 
     if (!(compare(x2, v2))) { 
        throw new java.lang.Exception("Exception test case 2 did not pass. x2= " + x2);} 
    ........................
    }
}

\end{lstlisting}
\end{table}

\section{Results}
This study presents a comprehensive evaluation of three language models—Gemini 1.5, Gemini 2.0, and LLaMA 7B—across 12 programming languages and six Indic languages using the CodeBERTScore as the primary metric. Overall, the Gemini models demonstrated superior performance compared to LLaMA 7B, with nuanced differences observed across the languages. The findings also provide interesting insights regarding language-specific challenges and strengths, particularly in less-resourced languages such as Sanskrit, and a remarkable consistency in Python code generation.

\subsection{Evaluation Results}
The investigation employed the CodeBERTScore to assess textual alignment and code generation capabilities across the diverse languages. Three models were evaluated:
\newline
\newline
\textbf{Gemini 1.5}: Achieved consistently high scores, ranging from 0.879 to 0.940. \textbf{Gemini 2.0}: Recorded scores between 0.875 and 0.946, outperforming Gemini 1.5 in select programming languages. \textbf{LLaMA 7B}: Although it maintained stable performance with scores from 0.794 to 0.870, it generally lagged behind the Gemini models.

\subsection{High alignment languages}
JavaScript, Python, and PHP recorded the highest CodeBERTScores. Both Gemini models typically exceeded a threshold of 0.930 in these languages. For instance, Gemini 1.5 excelled in JavaScript (with particularly high results for Hindi) while Gemini 2.0 exhibited excellence in PHP and Python—highlighting a model-dependent performance benefit in language-specific contexts.

\subsection{Moderate Alignment Languages}
Languages such as Java and Scala showed comparatively lower scores, with ranges of 0.875–0.907 for Java and 0.889–0.931 for Scala. These findings suggest that structural and syntactical nuances specific to these languages may impose additional challenges for current models.

\subsection{Superior Indic Language Performance}
Hindi, Urdu, and Punjabi achieved the highest alignment scores. Gemini 1.5 and 2.0 demonstrated robust performance across these languages, with some scores reaching as high as 0.939.

\subsection{Sanskrit in spotlight}
Sanskrit emerged as the most challenging Indic language, with scores varying from 0.794 to 0.917. This trend is likely attributable to the limited training data available for Sanskrit and its classical grammatical structure, which contrasts with the more contemporary usage patterns of the other Indic languages evaluated.

\section{Analysis and Discussion}
The performance analysis of the evaluated language models uncovers distinct strengths and opportunities for further enhancement across both programming and Indic languages. The Gemini models, in particular, consistently demonstrate superior semantic accuracy and code generation capabilities, as evidenced by their high CodeBERTScores. These models reliably preserve the underlying semantics across languages, which is critical for ensuring code correctness and functionality in diverse programming environments.
\newline
\newline
Notably, the analysis reveals that languages with relatively simpler syntactic structures, such as JavaScript and Python, achieve superior performance. This suggests that the current model architectures are well suited to manage less complex syntax, leading to higher alignment scores. In contrast, languages with more rigid and highly structured syntactic rules, such as Java and Scala, exhibit comparatively lower scores. This discrepancy indicates that further adaptation and fine-tuning may be necessary for models to more effectively handle the constraints imposed by these languages.
\newline
\newline
Through this work, we underscore the inherent shortcomings of the pass@k evaluation technique for generated code. Specifically, this method tends to under-represent the true capabilities of language models and, in particular, penalizes small language models, even when the generated code is functionally correct. This observation indicates a clear need for more robust, quantitative testing methods that accommodate the nuances of code produced by generative models. Future metrics should incorporate test cases that recognize acceptable variations in code structure and naming, thereby providing a more accurate reflection of a model's performance.
\newline
\newline
In addition to the programming languages, the evaluation across Indic languages presents important insights. While the Gemini models perform robustly in languages like Hindi, Urdu, and Punjabi, the results for Sanskrit highlight a critical area for improvement. The relatively lower scores in Sanskrit—attributable to its limited training resources and classical grammatical structure—underscore the need for enhanced fine-tuning and expansion of training datasets. Such efforts are crucial for improving semantic alignment and ensuring that multilingual models can generate accurate code in underrepresented languages.
\newline
\newline
These findings collectively reinforce the potential of multilingual code generation models while also emphasizing the necessity for continuous advancements. Enhancing code generation for low-resource languages will require strategic refinements in both model tuning and dataset augmentation. Future research should focus on tailored adaptations that address the syntactic and semantic complexities unique to underrepresented languages, thus contributing to broader linguistic and programming paradigm coverage essential for next-generation applications.

\section{Conclusion}
In this work, we presented IndicEval-XL, a multilingual code generation benchmark specifically designed for Indic languages. This benchmark aims to assess the capabilities of LLMs in generating code across diverse linguistic contexts. Our study highlights the challenges associated with cross-lingual code generation and points out significant limitations in current evaluation frameworks, such as the pass@k metric, which may fail to capture the true functional correctness of generated code. These observations underscore the need for more sophisticated evaluation methodologies and improved generative AI technologies. Future research should focus on developing robust metrics and models that enhance accessibility and performance across diverse linguistic and cultural settings, ensuring that technological advancements benefit a broader audience.

\section{Acknowledgments}
We would like to express our gratitude to the anonymous reviewers for their valuable feedback and insights, which have significantly contributed to refining this work. We also extend our thanks to the individuals who collaborated with us in validating the generated sample data for accuracy and consistency. This research was fully supported by Deutsche Telekom AG and Deutsche Telekom Digital Labs, and we deeply appreciate their support and resources.

\bibliographystyle{IEEEtran}
\bibliography{references}

\newpage
\newpage
\section{Model Performance Analysis}

This appendix presents a comprehensive evaluation of three language models (Gemini 1.5, Gemini 2.0, and LLaMA 7B) across 12 different programming languages. The analysis evaluates model's performance in understanding and processing natural language instructions in 6 different Indian languages: Hindi, Marathi, Punjabi, Sanskrit, Urdu, and Tamil.

\subsection{CodeBERT Results}
The following tables showcase the performance metrics obtained using CodeBERT for each programming language. 

\vspace{10pt}
\begin{table}[!ht]
    \centering
    \begin{tabular}{l c c c}
        \hline
        \textbf{Natural Language} & \textbf{Gemini 1.5} & \textbf{Gemini 2.0} & \textbf{LLaMA 7B} \\
        \hline
        Hindi       & 0.9394   & 0.9324   & 0.8487 \\
        Marathi     & 0.9311   & 0.9300   & 0.8249 \\
        Punjabi     & 0.9401   & 0.9315   & 0.8252 \\
        Sanskrit    & 0.9131   & 0.9190   & 0.8064 \\
        Urdu        & 0.9311   & 0.9339   & 0.8449 \\
        Tamil       & 0.9242   & 0.9245   & 0.8131 \\
        \hline
    \end{tabular}
\vspace{0.2cm} 
    \caption{Results (CodeBERT) of different models on Csharp across 6 natural languages.}
\end{table}
\vspace{10pt}

\begin{table}[!ht]
    \centering
    \begin{tabular}{l c c c}
        \hline
        \textbf{Natural Language} & \textbf{Gemini 1.5} & \textbf{Gemini 2.0} & \textbf{LLaMA 7B} \\
        \hline
        Hindi       & 0.9322   & 0.9294   & 0.8477 \\
        Marathi     & 0.9229   & 0.9212   & 0.8419 \\
        Punjabi     & 0.9313   & 0.9292   & 0.8432 \\
        Sanskrit    & 0.9114   & 0.9083   & 0.8319 \\
        Urdu        & 0.9241   & 0.9290   & 0.8508 \\
        Tamil       & 0.9159   & 0.9271   & 0.8258 \\
        \hline
    \end{tabular}
\vspace{0.2cm} 
    \caption{Results (CodeBERT) of different models on Go across 6 natural languages.}
\end{table}
\vspace{10pt}
\begin{table}[!ht]
    \centering
    \begin{tabular}{l c c c}
        \hline
        \textbf{Natural Language} & \textbf{Gemini 1.5} & \textbf{Gemini 2.0} & \textbf{LLaMA 7B} \\
        \hline
        Hindi       & 0.9035   & 0.8959   & 0.8498 \\
        Marathi     & 0.9023   & 0.8887   & 0.8254 \\
        Punjabi     & 0.9076   & 0.8902   & 0.8463 \\
        Sanskrit    & 0.8792   & 0.8758   & 0.8189 \\
        Urdu        & 0.9057   & 0.8878   & 0.8435 \\
        Tamil       & 0.8972   & 0.8837   & 0.8184 \\
        \hline
    \end{tabular}
\vspace{0.2cm} 
    \caption{Results (CodeBERT) of different models on Java across 6 natural languages.}
\end{table}
\vspace{10pt}
\begin{table}[!ht]
    \centering
    \begin{tabular}{l c c c}
        \hline
        \textbf{Natural Language} & \textbf{Gemini 1.5} & \textbf{Gemini 2.0} & \textbf{LLaMA 7B} \\
        \hline
        Hindi       & 0.9364   & 0.9205   & 0.8660 \\
        Marathi     & 0.9306   & 0.9172   & 0.8397 \\
        Punjabi     & 0.9349   & 0.9239   & 0.8488 \\
        Sanskrit    & 0.9134   & 0.9008   & 0.8203 \\
        Urdu        & 0.9369   & 0.9291   & 0.8544 \\
        Tamil       & 0.9271   & 0.9005   & 0.8177 \\
        \hline
    \end{tabular}
\vspace{0.2cm} 
    \caption{Results (CodeBERT) of different models on Javascript across 6 natural languages.}
\end{table}
\vspace{10pt}
\begin{table}[!ht]
    \centering
    \begin{tabular}{l c c c}
        \hline
        \textbf{Natural Language} & \textbf{Gemini 1.5} & \textbf{Gemini 2.0} & \textbf{LLaMA 7B} \\
        \hline
        Hindi       & 0.9198   & 0.9326   & 0.8541 \\
        Marathi     & 0.9229   & 0.9229   & 0.8297 \\
        Punjabi     & 0.9250   & 0.9355   & 0.8434 \\
        Sanskrit    & 0.9065   & 0.9089   & 0.8069 \\
        Urdu        & 0.9294   & 0.9321   & 0.8445 \\
        Tamil       & 0.9180   & 0.9119   & 0.8017 \\
        \hline
    \end{tabular}
\vspace{0.2cm} 
    \caption{Results (CodeBERT) of different models on Kotlin across 6 natural languages.}
\end{table}
\vspace{10pt}
\begin{table}[!ht]
    \centering
    \begin{tabular}{l c c c}
        \hline
        \textbf{Natural Language} & \textbf{Gemini 1.5} & \textbf{Gemini 2.0} & \textbf{LLaMA 7B} \\
        \hline
        Hindi       & 0.9283   & 0.9288   & 0.8348 \\
        Marathi     & 0.9230   & 0.9146   & 0.8068 \\
        Punjabi     & 0.9313   & 0.9228   & 0.8252 \\
        Sanskrit    & 0.9038   & 0.8933   & 0.7947 \\
        Urdu        & 0.9270   & 0.9271   & 0.8382 \\
        Tamil       & 0.9180   & 0.9058   & 0.8015 \\
        \hline
    \end{tabular}
\vspace{0.2cm} 
    \caption{Results (CodeBERT) of different models on Perl across 6 natural languages.}
\end{table}
\vspace{10pt}
\begin{table}[!ht]
    \centering
    \begin{tabular}{l c c c}
        \hline
        \textbf{Natural Language} & \textbf{Gemini 1.5} & \textbf{Gemini 2.0} & \textbf{LLaMA 7B} \\
        \hline
        Hindi       & 0.9354   & 0.9361   & 0.8421 \\
        Marathi     & 0.9234   & 0.9314   & 0.8183 \\
        Punjabi     & 0.9353   & 0.9468   & 0.8230 \\
        Sanskrit    & 0.9128   & 0.9228   & 0.7980 \\
        Urdu        & 0.9310   & 0.9465   & 0.8449 \\
        Tamil       & 0.9181   & 0.9284   & 0.8094 \\
        \hline
    \end{tabular}
\vspace{0.2cm} 
    \caption{Results (CodeBERT) of different models on Php across 6 natural languages.}
\end{table}
\vspace{10pt}
\begin{table}[!ht]
    \centering
    \begin{tabular}{l c c c}
        \hline
        \textbf{Natural Language} & \textbf{Gemini 1.5} & \textbf{Gemini 2.0} & \textbf{LLaMA 7B} \\
        \hline
        Hindi       & 0.92099   & 0.93882   & 0.87019 \\
        Marathi     & 0.90904   & 0.92319   & 0.84056 \\
        Punjabi     & 0.92441   & 0.93287   & 0.84806 \\
        Sanskrit    & 0.91320   & 0.90392   & 0.81718 \\
        Urdu        & 0.92110   & 0.93905   & 0.86218 \\
        Tamil       & 0.91303   & 0.89991   & 0.84271 \\
        \hline
    \end{tabular}
\vspace{0.2cm} 
    \caption{Results (CodeBERT) of different models on Python across 6 natural languages.}
\end{table}
\vspace{10pt}
\begin{table}[!ht]
    \centering
    \begin{tabular}{l c c c}
        \hline
        \textbf{Natural Language} & \textbf{Gemini 1.5} & \textbf{Gemini 2.0} & \textbf{LLaMA 7B} \\
        \hline
        Hindi       & 0.9316   & 0.9362   & 0.8431 \\
        Marathi     & 0.9195   & 0.9319   & 0.8294 \\
        Punjabi     & 0.9307   & 0.9420   & 0.8379 \\
        Sanskrit    & 0.9109   & 0.9104   & 0.7989 \\
        Urdu        & 0.9262   & 0.9415   & 0.8456 \\
        Tamil       & 0.9182   & 0.9181   & 0.8085 \\
        \hline
    \end{tabular}
\vspace{0.2cm} 
    \caption{Results (CodeBERT) of different models on Ruby across 6 natural languages.}
\end{table}
\vspace{10pt}
\begin{table}[H]
    \setlength{\abovecaptionskip}{0pt}
    \setlength{\belowcaptionskip}{0pt}
    \centering
    \small 
    \renewcommand{\arraystretch}{0.9}
    \begin{tabular}{l c c c}
        \hline
        \textbf{Natural Language} & \textbf{Gemini 1.5} & \textbf{Gemini 2.0} & \textbf{LLaMA 7B} \\
        \hline
        Hindi       & 0.9104   & 0.9184   & 0.8349 \\
        Marathi     & 0.9008   & 0.9161   & 0.8081 \\
        Punjabi     & 0.9023   & 0.9283   & 0.8252 \\
        Sanskrit    & 0.8895   & 0.8964   & 0.8078 \\
        Urdu        & 0.9160   & 0.9315   & 0.8370 \\
        Tamil       & 0.8912   & 0.9194   & 0.8064 \\
        \hline
    \end{tabular}
\vspace{0.2cm} 
    \caption{Results (CodeBERT) of different models on Scala across 6 natural languages.}
\end{table}

\begin{table}[H]
    \vspace {-5pt}
    \setlength{\abovecaptionskip}{0pt}
    \setlength{\belowcaptionskip}{0pt}
    \centering
    \small 
    \renewcommand{\arraystretch}{0.9}
    \begin{tabular}{l c c c}
        \hline
        \textbf{Natural Language} & \textbf{Gemini 1.5} & \textbf{Gemini 2.0} & \textbf{LLaMA 7B} \\
        \hline
        Hindi       & 0.9304   & 0.9312   & 0.8532 \\
        Marathi     & 0.9274   & 0.9225   & 0.8321 \\
        Punjabi     & 0.9345   & 0.9287   & 0.8396 \\
        Sanskrit    & 0.9100   & 0.9092   & 0.8122 \\
        Urdu        & 0.9288   & 0.9313   & 0.8473 \\
        Tamil       & 0.9166   & 0.9186   & 0.8143 \\
        \hline
    \end{tabular}
\vspace{0.2cm} 
    \caption{Results (CodeBERT) of different models on Swift across 6 natural languages.}
\end{table}

\begin{table}[!ht]
    \vspace {-5pt}
    \setlength{\abovecaptionskip}{0pt}
    \setlength{\belowcaptionskip}{0pt}
    \centering
    \small 
    \renewcommand{\arraystretch}{0.9}
    \begin{tabular}{l c c c}
        \hline
        \textbf{Natural Language} & \textbf{Gemini 1.5} & \textbf{Gemini 2.0} & \textbf{LLaMA 7B} \\
        \hline
        Hindi       & 0.9315   & 0.9352   & 0.8632 \\
        Marathi     & 0.9268   & 0.9275   & 0.8338 \\
        Punjabi     & 0.9371   & 0.9441   & 0.8416 \\
        Sanskrit    & 0.9171   & 0.9124   & 0.8098 \\
        Urdu        & 0.9358   & 0.9440   & 0.8494 \\
        Tamil       & 0.9259   & 0.9231   & 0.8194 \\
        \hline
    \end{tabular}
\vspace{0.2cm} 
    \caption{Results (CodeBERT) of different models on Typescript across 6 natural languages.}
\end{table}
\vspace{10pt}
\clearpage

\section{BERTScore Model Analysis}
This section presents the performance analysis using BERTScore, which is particularly effective in assessing the quality of generated text across multiple languages.

\subsection{BERTscore Results}

\vspace{10pt}

\begin{table}[!ht]
    \centering
    \begin{tabular}{l c c c}
        \hline
        \textbf{Natural Language} & \textbf{Gemini 1.5} & \textbf{Gemini 2.0} & \textbf{LLaMA 7B} \\
        \hline
        Tamil            & 0.7960   & 0.7984   & 0.4805 \\
        Hindi            & 0.8343   & 0.8199   & 0.5527 \\
        Urdu             & 0.8112   & 0.8269   & 0.5420 \\
        Sanskrit         & 0.7703   & 0.7922   & 0.4609 \\
        Punjabi         & 0.8334   & 0.8217   & 0.4917 \\
        Marathi          & 0.8184   & 0.8117   & 0.5087 \\
        \hline
    \end{tabular}
\vspace{0.2cm} 
    \caption{Results (Bertscore) of different models on Csharp across 6 natural languages.}
\end{table}
\vspace{10pt}
\begin{table}[!ht]
    \centering
    \begin{tabular}{l c c c}
        \hline
        \textbf{Natural Language} & \textbf{Gemini 1.5} & \textbf{Gemini 2.0} & \textbf{LLaMA 7B} \\
        \hline
        Tamil           & 0.7176   & 0.7120   & 0.4083 \\
        Hindi           & 0.7634   & 0.7328   & 0.4694 \\
        Urdu            & 0.7351   & 0.7231   & 0.4635 \\
        Sanskrit        & 0.6987   & 0.6658   & 0.4059 \\
        Punjabi         & 0.7617   & 0.7242   & 0.4602 \\
        Marathi         & 0.7343   & 0.7017   & 0.4470 \\
        \hline
    \end{tabular}
\vspace{0.2cm} 
    \caption{Results (Bertscore) of different models on Go across 6 natural languages.}
\end{table}
\vspace{10pt}
\begin{table}[!ht]
    \centering
    \begin{tabular}{l c c c}
        \hline
        \textbf{Natural Language} & \textbf{Gemini 1.5} & \textbf{Gemini 2.0} & \textbf{LLaMA 7B} \\
        \hline
        Tamil           & 0.7399   & 0.7223   & 0.5611 \\
        Hindi           & 0.7459   & 0.7502   & 0.6312 \\
        Urdu            & 0.7532   & 0.7311   & 0.6188 \\
        Sanskrit        & 0.6983   & 0.6999   & 0.5650 \\
        Punjabi         & 0.7593   & 0.7320   & 0.6206 \\
        Marathi         & 0.7480   & 0.7288   & 0.5764 \\
        \hline
    \end{tabular}
\vspace{0.2cm} 
    \caption{Results (Bertscore) of different models on Java across 6 natural languages.}
\end{table}
\vspace{10pt}
\begin{table}[!ht]
    \centering
    \begin{tabular}{l c c c}
       \hline
       \textbf{Natural Language} & \textbf{Gemini 1.5} & \textbf{Gemini 2.0} & \textbf{LLaMA 7B} \\
       \hline
       Tamil           & 0.7522   & 0.6821   & 0.4173 \\
       Hindi           & 0.7741   & 0.7489   & 0.5630 \\
       Urdu            & 0.7850   & 0.7722   & 0.5262 \\
       Sanskrit        & 0.7003   & 0.6833   & 0.4364 \\
       Punjabi         & 0.7700   & 0.7447   & 0.5046 \\
       Marathi         & 0.7694   & 0.7371   & 0.4776 \\
       \hline
    \end{tabular}
\vspace{0.2cm} 
    \caption{Results (Bertscore) of different models on Javascript across 6 natural languages.}
\end{table}
\vspace{10pt}
\begin{table}[!ht]
    \centering
    \begin{tabular}{l c c c}
       \hline
       \textbf{Natural Language} & \textbf{Gemini 1.5} & \textbf{Gemini 2.0} & \textbf{LLaMA 7B} \\
       \hline
       Tamil           & 0.7208   & 0.7275   & 0.4124 \\
       Hindi           & 0.7343   & 0.7835   & 0.5565 \\
       Urdu            & 0.7614   & 0.7863   & 0.5147 \\
       Sanskrit        & 0.7039   & 0.7187   & 0.4280 \\
       Punjabi         & 0.7522   & 0.7882   & 0.5101 \\
       Marathi         & 0.7497   & 0.7645   & 0.4894 \\
       \hline
    \end{tabular}
\vspace{0.2cm} 
    \caption{Results (Bertscore) of different models on Kotlin across 6 natural languages.}
\end{table}
\vspace{10pt}
\begin{table}[!ht]
    \centering
    \begin{tabular}{l c c c}
        \hline
        \textbf{Natural Language} & \textbf{Gemini 1.5} & \textbf{Gemini 2.0} & \textbf{LLaMA 7B} \\
        \hline
        Tamil           & 0.7195   & 0.6838   & 0.4034 \\
        Hindi           & 0.7468   & 0.7519   & 0.4929 \\
        Urdu            & 0.7429   & 0.7464   & 0.4979 \\
        Sanskrit        & 0.6774   & 0.6663   & 0.4150 \\
        Punjabi         & 0.7588   & 0.7340   & 0.4622 \\
        Marathi         & 0.7357   & 0.7188   & 0.4294 \\
        \hline
    \end{tabular}
\vspace{0.2cm} 
    \caption{Results (Bertscore) of different models on Perl across 6 natural languages.}
\end{table}
\vspace{10pt}
\begin{table}[!ht]
    \centering
    \begin{tabular}{l c c c}
        \hline
        \textbf{Natural Language} & \textbf{Gemini 1.5} & \textbf{Gemini 2.0} & \textbf{LLaMA 7B} \\
        \hline
        Tamil           & 0.7566   & 0.7648   & 0.4352 \\
        Hindi           & 0.8114   & 0.8020   & 0.5289 \\
        Urdu            & 0.7918   & 0.8261   & 0.5302 \\
        Sanskrit        & 0.7537   & 0.7602   & 0.4208 \\
        Punjabi         & 0.8010   & 0.8245   & 0.4654 \\
        Marathi         & 0.7696   & 0.7904   & 0.4670 \\
        \hline
    \end{tabular}
\vspace{0.2cm} 
    \caption{Results (Bertscore) of different models on PHP across 6 natural languages.}
\end{table}
\vspace{10pt}
\begin{table}[!ht]
    \centering
    \begin{tabular}{l c c c}
       \hline
       \textbf{Natural Language} & \textbf{Gemini 1.5} & \textbf{Gemini 2.0} & \textbf{LLaMA 7B} \\
       \hline
       Tamil           & 0.6975   & 0.6558   & 0.4600 \\
       Hindi           & 0.7255   & 0.7832   & 0.5480 \\
       Urdu            & 0.7313   & 0.7850   & 0.5266 \\
       Sanskrit        & 0.7024   & 0.6745   & 0.3971 \\
       Punjabi         & 0.7287   & 0.7618   & 0.4568 \\
       Marathi         & 0.6883   & 0.7357   & 0.4718 \\
       \hline
    \end{tabular}
\vspace{0.2cm} 
    \caption{Results (Bertscore) of different models on Python across 6 natural languages.}
\end{table}
\vspace{10pt}
\begin{table}[!ht]
    \centering
    \begin{tabular}{l c c c}
       \hline
       \textbf{Natural Language} & \textbf{Gemini 1.5} & \textbf{Gemini 2.0} & \textbf{LLaMA 7B} \\
       \hline
       Tamil           & 0.6863   & 0.7027   & 0.3204 \\
       Hindi           & 0.7333   & 0.7546   & 0.4513 \\
       Urdu            & 0.7179   & 0.7663   & 0.4484 \\
       Sanskrit        & 0.6508   & 0.6675   & 0.3069 \\
       Punjabi         & 0.7258   & 0.7741   & 0.4073 \\
       Marathi         & 0.6891   & 0.7464   & 0.3920 \\
       \hline
    \end{tabular}
\vspace{0.2cm} 
    \caption{Results (Bertscore) of different models on Ruby across 6 natural languages.}
\end{table}
\vspace{10pt}
\begin{table}[!ht]
    \centering
    \begin{tabular}{l c c c}
       \hline
       \textbf{Natural Language} & \textbf{Gemini 1.5} & \textbf{Gemini 2.0} & \textbf{LLaMA 7B} \\
       \hline
       Tamil           & 0.6540   & 0.7482   & 0.3731 \\
       Hindi           & 0.7036   & 0.7423   & 0.4626 \\
       Urdu            & 0.7202   & 0.7740   & 0.4542 \\
       Sanskrit        & 0.6443   & 0.6774   & 0.3892 \\
       Punjabi         & 0.6815   & 0.7679   & 0.4179 \\
       Marathi         & 0.6877   & 0.7407   & 0.3814 \\
       \hline
    \end{tabular}
\vspace{0.2cm} 
    \caption{Results (Bertscore) of different models on Scala across 6 natural languages.}
\end{table}

\begin{table}[H]
    \setlength{\abovecaptionskip}{0pt}
    \setlength{\belowcaptionskip}{0pt}
    \centering
    \small 
    \renewcommand{\arraystretch}{0.9} 
    \begin{tabular}{l c c c}
       \hline
       \textbf{Natural Language} & \textbf{Gemini 1.5} & \textbf{Gemini 2.0} & \textbf{LLaMA 7B} \\
       \hline
       Tamil           & 0.7406   & 0.7433   & 0.4378 \\
       Hindi           & 0.7772   & 0.7860   & 0.5475 \\
       Urdu            & 0.7683   & 0.7845   & 0.5386 \\
       Sanskrit        & 0.7160   & 0.7225   & 0.4471 \\
       Punjabi         & 0.7837   & 0.7774   & 0.5013 \\
       Marathi         & 0.7672   & 0.7605   & 0.5032 \\
       \hline
    \end{tabular}
    \caption{Results (Bertscore) of different models on Swift across 6 natural languages.}
\end{table}

\begin{table}[H]
    \vspace{-5pt} 
    \setlength{\abovecaptionskip}{0pt}
    \setlength{\belowcaptionskip}{0pt}
    \centering
    \small 
    \renewcommand{\arraystretch}{0.9} 
    \begin{tabular}{l c c c}
       \hline
       \textbf{Natural Language} & \textbf{Gemini 1.5} & \textbf{Gemini 2.0} & \textbf{LLaMA 7B} \\
       \hline
       Tamil           & 0.7367   & 0.7441   & 0.4254 \\
       Hindi           & 0.7655   & 0.7756   & 0.5507 \\
       Urdu            & 0.7776   & 0.7978   & 0.5154 \\
       Sanskrit        & 0.7161   & 0.6996   & 0.4228 \\
       Punjabi         & 0.7748   & 0.7947   & 0.4854 \\
       Marathi         & 0.7439   & 0.7560   & 0.4690 \\
       \hline
    \end{tabular}
    \caption{Results (Bertscore) of different models on TypeScript across 6 natural languages.}
\end{table}
\clearpage

\newpage
\begin{center}
\vspace{0.5cm} 
\begin{figure}[H]
    \centering
    \includegraphics[width=\textwidth]{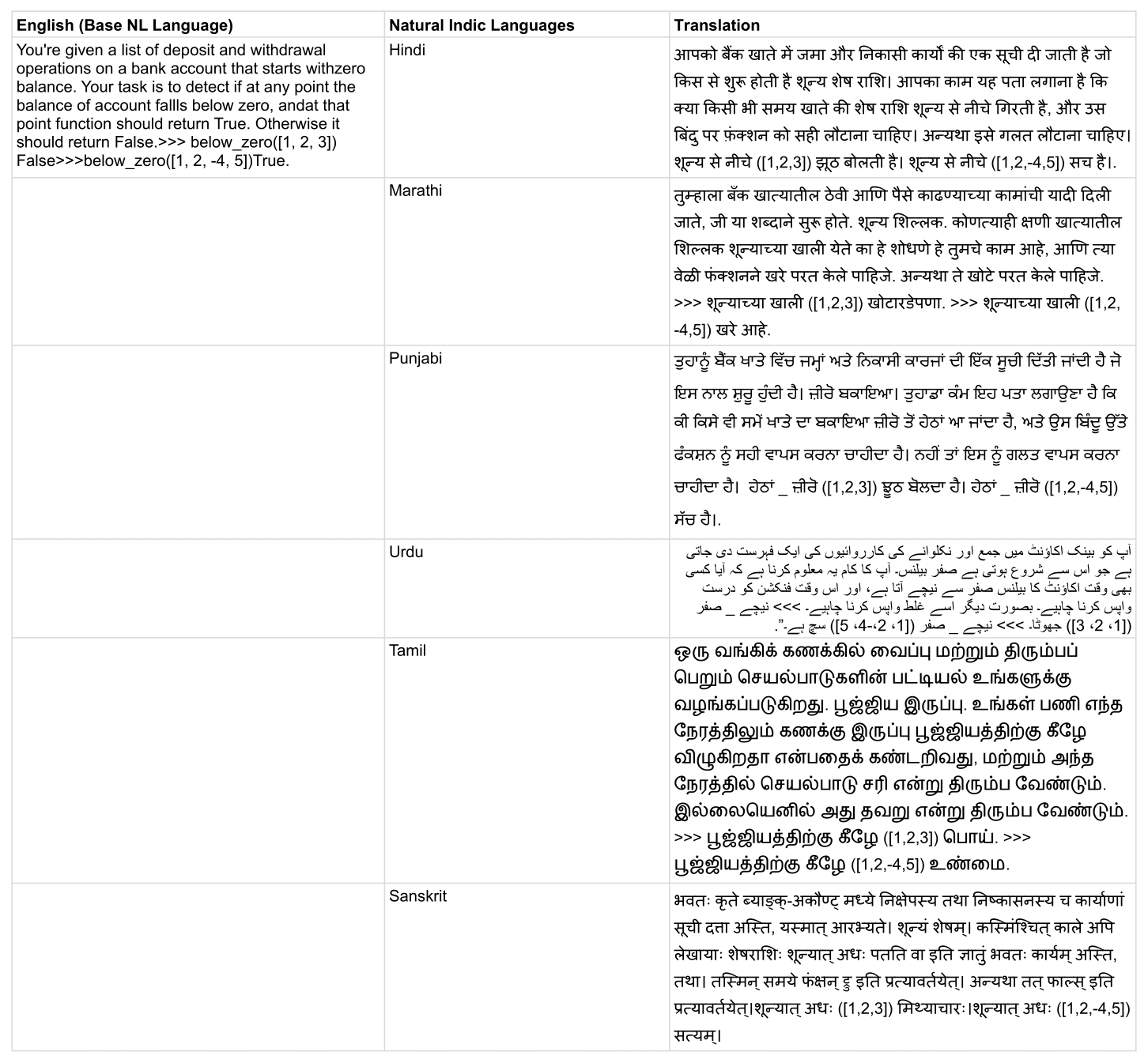} 
    \caption{Translation examples across different natural languages for TypeScript.}
    \label{fig:translation_examples}
\end{figure}
\end{center}

\end{document}